\documentclass[
  a4paper
]{cup-journal}

\usepackage{amsmath}
\usepackage{hyperref}
\usepackage{booktabs}
\usepackage{xcolor}
\usepackage{tabularx}
\usepackage{graphicx}
\usepackage{array}
\usepackage{enumitem}
\usepackage[nopatch]{microtype}
\usepackage{balance}
\usepackage{makecell}
\usepackage{multicol}
\usepackage{mhchem}
\usepackage{graphicx}
\usepackage{caption}
\usepackage{subcaption}
\usepackage{float}
\usepackage{indentfirst}
\usepackage{cite}
\RequirePackage{geometry}
\usepackage{fixltx2e}
\usepackage{braket}

\title{Leveraging Machine Learning to Overcome Limitations in Quantum Algorithms}

\author{Laia Coronas Sala$^{1}$, Parfait Atchade-Adelemou$^{2,3}$}


\begin{document} 

\maketitle
\noindent
$^{1}$ University of Barcelona, Biomedical Engineering, Barcelona, Spain. Email: lclaiacoronas@gmail.com\\
$^{2}$ Lighthouse Disruptive Innovation Group S.L, Barcelona, Spain. Email: parfait.atchade@lighthouse-dig.com
\\$^{3}$ MIT Media Lab, City Science, Boston, MA, USA. Email: parfait@mit.edu

\noindent
\begin{abstract}
\noindent \textbf{Abstract}

\noindent 
Quantum Computing (QC) offers outstanding potential for molecular characterization and drug discovery, particularly in solving complex properties like the Ground State Energy (GSE) of biomolecules. However, QC is constrained by computational noise, scalability, and system complexity. This work presents a hybrid framework that combines Machine Learning (ML) techniques with quantum algorithms, namely the Variational Quantum Eigensolver (VQE), Hartree-Fock (HF), and Quantum Phase Estimation (QPE)—to improve GSE predictions when computing large molecules.
We curated three datasets—chemical descriptors, Coulomb matrices, and a hybrid combination—using molecular features from PubChem. These datasets were used to train XGBoost (XGB), Random Forest (RF), and LightGBM (LGBM) models. Results show that XGB achieved the lowest Relative Error (RE) of 4.41 ± 11.18\% on chemical descriptors, outperforming RF (5.56 ± 11.66\%) and LGBM (5.32 ± 12.87\%). The HF method delivered exceptional precision for small molecules (0.44 ± 0.66\% RE), while a near-linear correlation between GSE and molecular electron count provided predictive shortcuts.
This study demonstrates that integrating QC and ML enhances scalability for molecular energy predictions. The proposed approach lays the foundation for overcoming QC’s limitations and scaling molecular simulations to larger systems.

\end{abstract}
\begin{multicols}{2}

\section{INTRODUCTION}
Accurate protein and drug molecule characterization has driven the development of diverse computational and experimental techniques \cite{Yu2017}. Since John Dalton’s 19\textsuperscript{th}-century atomic theory, it is known that molecules comprise atoms made of electrons, protons, and neutrons \cite{grossman2021}. This raises a critical question: \textit{Can deeper insights into subatomic behavior enhance molecular characterization?} \\
While modern tools like AlphaFold, Rosetta, and I-TASSER have revolutionized the prediction of three-dimensional structures \cite{jumper2021alphafold, leaver2011rosetta, yang2015itasser}, they rely on approximations that can limit accuracy, increasing costs and timelines in drug development \cite{hughes2011principles, yu2017computer, shinn2019high, lin2020review}. Quantum mechanics, by contrast, explores particle behavior at subatomic scales, capturing interactions and energy states with unparalleled precision. Unlike homology-dependent classical methods, quantum chemistry is grounded in physical principles of energy and entropy, making it ideal for high-precision conformational predictions \cite{vanmourik2004firstprinciples, bryce2020next}.
Key quantum properties such as superposition enable quantum computers to process multiple states simultaneously, vastly outperforming classical systems in complex computations \cite{nielsen2010quantum}. 

Quantum mechanics models biological systems via wave functions, solutions to the Schrödinger equation \eqref{eq:shrodinger}, which predict electron behavior and molecular properties. 

\begin{equation}
i\hbar \frac{\partial}{\partial t} |\psi\rangle = \hat{H} |\psi\rangle,
\label{eq:shrodinger}
\end{equation}

The Hamiltonian operator $\hat{H}$, Eq:\eqref{eq:BornOp_app}, represents system energy, simplified by the Born-Oppenheimer approximation \cite{li2023}.

\begin{equation}
\hat{H} = - \sum_{i=1}^N \frac{1}{2} \nabla_i^2 - \sum_{i=1}^N \sum_{A=1}^M \frac{Z_A}{r_{iA}} + \sum_{j>i} \frac{1}{r_{ij}},
\label{eq:BornOp_app}
\end{equation}

Where \(\hat{H}\) represents the Hamiltonian operator, \(N\) is the total number of electrons in the system, \(\nabla_i^2\) is the Laplacian operator acting on the \(i\)-th electron, \(M\) is the total number of nuclei in the system, \(Z_A\) is the atomic number of the \(A\)-th nucleus representing the charge of the nucleus, \(r_{iA}\) is the distance between the \(i\)-th electron and the \(A\)-th nucleus, and \(r_{ij}\) is the distance between the \(i\)-th and \(j\)-th electrons. Solving the time-independent Schrödinger equation \eqref{eq:timeind} yields the eigenvalues $E$, representing allowed energy states:

\begin{equation}
\hat{H} |\psi\rangle = E |\psi\rangle,
\label{eq:timeind}
\end{equation}

The lowest eigenvalue corresponds to the Ground State Energy (GSE), the system’s most stable configuration \cite{helmenstine}. This state underpins molecular stability, reactivity, and interaction dynamics, informing reaction pathways and drug binding processes \cite{britt2004}. However, solving such equations for relatively small systems is already very challenging since it involves solving high search space equations. Therefore, developing efficient and scalable procedures to compute the GSE for large molecules remains an open challenge.

\subsection{Motivation}

The motivation for this work stems from the challenges of finding exact solutions for quantum equations in biological systems with over 30 electrons due to the computational complexity of quantum simulations \cite{blunt2022}. While awaiting advancements in hardware and microelectronics, this project explores strategies to mitigate qubit limitations by integrating quantum methods, datasets, and ML for analyzing complex biomolecules. 

To test this approach, we focused on computing the ground-state energies of the 20 essential amino acids found in the human body due to their profound biological relevance \cite{berg2002biochemistry}. As the building blocks of proteins, these amino acids play a central role in numerous biochemical processes, making them ideal candidates for exploring molecular interactions and dynamics while highlighting the practical implications of this research. Furthermore, this strategy aims to provide the necessary tools for scaling up to larger molecular systems, such as small proteins.

With this ultimate objective, we set the following pipeline:

\begin{itemize} \item Develop a pipeline to assess the GSEs, examining different Hamiltonian encoders (Jordan-Wigner and Bravyi-Kitaev \cite{Veyrac2024, tranter2015bravyi, seeley2012bravyi}) and quantum strategies (Quantum Phase Estimation [QPE] \cite{Kang2022}, Variational Quantum Eigensolver [VQE] \cite{kirby2021, peruzzo2014variational,adelomou2020usingparameterizedquantumcircuit,10.1007/978-3-030-61705-9_21}, and the Hartree-Fock Approximation [HF] \cite{Crisostomo2022, slater1928self, slater1930note}). \item Create robust training datasets incorporating various molecular descriptors to train ML models and identify feature combinations that yield optimal accuracies. \item Build ML models trained on these datasets to predict GSEs for larger molecules that cannot be directly simulated using QC, focusing on the $20$ essential amino acids. \end{itemize}

\section{RELATED WORK}
The first step before any simulation can begin is encoding the Hamiltonian. For that, the Jordan-Wigner transformation has shown efficacy in calculating the chemical properties of small molecules \cite{Li2022, Veyrac2024}. However, its long sequences of ZZ operations can significantly increase resource demands on quantum hardware \cite{barenco1995elementary}. The Bravyi-Kitaev transformation offers an alternative that optimizes qubit usage and enhances computational efficiency while maintaining accuracy \cite{seeley2012, setia2018bravyi, tranter2018comparison}. Once encoded, quantum strategies like QPE and VQE have been effective for extracting GSEs \cite{atchadeadelomou2023efficient,atchadeadelomou2022quantumalgorithmssolvinghard, alonsolinaje2021evaquantumexponentialvalue}. 

These quantum methods have been applied to small, computationally feasible molecules, however, for larger molecules, Machine learning (ML) has emerged as an alternative to extrapolate quantum properties to more complex systems that cannot be easily simulated. For example, Dominic \textit{et al.} and Lauv \textit{et al.} demonstrated promising results using ML models for generating drug-like molecules \cite{Martinelli2022, Patel2020}, supported by similar findings from other studies \cite{Vamathevan2019, Gupta2021, muller2022}. 

Recognizing the importance of robust training datasets, efforts have been made to construct comprehensive datasets. For instance, QM7-X, containing $4.2$ million small organic molecules, has proven valuable for predicting ground state properties \cite{hoja2021}. Predecessors like QM8 and QM9 have also contributed significantly \cite{ramakrishnan2014, ramakrishnan2015, ruddigkeit2012}, alongside QMugs, which focuses on drug-like molecules \cite{issert2022}. However, the need for larger, more diverse datasets persists, with some advocating for the inclusion of reoriented molecules to increase training samples and enhance accuracy \cite{hansen2013}.

Despite these advancements, current strategies for computing quantum properties in biological systems lack sufficient accuracy when applied to larger and more complex molecules. Additionally, existing training datasets often emphasize limited molecular features, such as three-dimensional coordinates and Coulomb matrices, while overlooking other important molecular characteristics. Therefore, this project aims to develop an effective strategy for calculating GSEs for large systems without directly solving quantum equations. Furthermore, given their pivotal role in protein structures and drug discovery, we focus on the 20 essential amino acids that constitute all of our proteins.

\section{BACKGROUND} 

\subsection{Quantum Chemistry}
Efficient encoding of the Hamiltonian is essential for the drugs discovery process, enabling quantum algorithms to compute each fragment's properties through different quantum algorithms such as the Hartree-Fock (HF) Approximation \cite{Crisostomo2022, slater1928self, slater1930note}, Variational Quantum Eigensolver (VQE) \cite{kandala2017hardware}, and Quantum Phase Estimation (QPE) \cite{QPE}.

\subsubsection{Hamiltonian Encoding}
Computing  GSEs requires encoding the Hamiltonian of molecular fragments in a form suitable for quantum simulations, as represented in equation \eqref{eq:BornOp_app}. Two primary methods for encoding are the Jordan-Wigner transformation \cite{jordan1928paulische, tranter2015bravyi} and the Bravyi-Kitaev transformation \cite{bravyi2002fermionic}, which efficiently map fermionic states to qubit representations for quantum computation. To translate particle systems into qubits, Pauli operators $X$, $Y$, and $Z$ are used to create and remove particles, as shown in equations \eqref{eq:creator_} and \eqref{eq:creator}.

\begin{equation}
Q^\dagger_j = \frac{1}{2} (X_j - i Y_j),
\label{eq:creator_}
\end{equation}

\begin{equation}
Q_j = \frac{1}{2} (X_j + i Y_j),
\label{eq:creator}
\end{equation}

Where \(X_j\) and \(Y_j\) are Pauli operators applied to the \(j\)-th qubit, \(i\) is the imaginary unit, \(Q_j^{\dagger}\) is the complex conjugate of \(Q_j\), \(X_j\) and \(Y_j\) are real variables associated with the \(j\)-th component.

To achieve this, a sequence of $Z$ operators is added, leading to the final Jordan-Wigner representation in equations \eqref{eq:creator_1} and \eqref{eq:descructor}:

\begin{equation}
a^\dagger_j = \frac{1}{2} (X_j - i Y_j) \otimes \bigotimes_{k<j} Z_k,
\label{eq:creator_1},
\end{equation}

\begin{equation}
a_j = \frac{1}{2} (X_j + i Y_j) \otimes \bigotimes_{k<j} Z_k,  
\label{eq:descructor}
\end{equation}

Where \(a_j^{\dagger}\) is the complex conjugate of \(a_j\), \(X_j\) and \(Y_j\) are real variables associated with the \(j\)-th component, \(i\) is the imaginary unit, and \(\bigotimes_{k<j} Z_k\) denotes the tensor product of \(Z_k\) operators for \(k\) less than \(j\). 

Once correctly encoded, the systems can be handled by a quantum algorithms that may be able to solve their ground states.

\subsubsection{Variational Quantum Eigensolver}

The Variational Quantum Eigensolver (VQE) \cite{kandala2017hardware} is a key quantum algorithm for finding the GSE of a quantum system given a molecular Hamiltonian and a parameterized circuit. VQE begins by preparing a quantum state using a parameterized circuit, known as an ansatz, derived from the system's fermionic representation mapped through Jordan-Wigner or Bravyi-Kitaev encodings \cite{atchadeadelomou2022quantum}. An initial state, often the Hartree-Fock state, is defined, and the circuit runs iteratively to collect energy measurements. These measurements are input into a classical optimization algorithm that adjusts the circuit parameters to minimize the energy. The state preparation, measurement, and optimization process repeats until the energy converges to a minimum, indicating the GSE of the system \cite{kirby2021}.

\subsubsection{Quantum Phase Estimation}

Quantum Phase Estimation (QPE) \cite{QPE} is a quantum algorithm used to estimate the eigenvalues of an eigenvector of a unitary operator, such as the molecular Hamiltonian of a fragment. By converting the Hamiltonian into a unitary operator through trotterization \cite{somma2016trotter}, QPE can determine its eigenvalues, including the GSE. The equation \eqref{eq:QPE} expresses the described feature.

\begin{equation}
U \ket{\psi} = e^{i\phi} \ket{\psi},
\label{eq:QPE}
\end{equation}

Where \(U\) is a unitary operator, \(|\psi\rangle\) is a quantum state, \(e^{i\phi}\) is a phase factor, \(i\) is the imaginary unit, and \(\phi\) is a real number representing the phase.

The process begins by preparing a quantum state that is a superposition of the eigenstates of \(U\). The exponential term \(e^{i\phi}\) can be rewritten as \(e^{2\pi i \theta}\), where \(\theta\) represents the phase in binary form. The next step involves applying the Quantum Fourier Transform (QFT) \cite{preskill1998lecture} and controlling unitary operations to encode eigenvalue. Finally, the inverse QFT is applied to measure the qubits and estimate the phase value, resulting in a set of eigenvalues, including the molecular system's GSE.
 
\subsubsection{Hartree-Fock Approximation}

The Hartree-Fock (HF) \cite{hartree1928wave} method is a computational technique used to determine the electronic structure of atoms and molecules. It employs a self-consistent field approach, where each electron is treated as moving within an average field created by all other electrons \cite{hartree1928wave,slater1928self,slater1930note}. The method iteratively refines the equations until self-consistency is achieved, ensuring that the input and output electron densities converge. The HF equations are derived from the variational principle, which seeks to minimize the system's total energy with respect to single-electron wave functions or orbitals. To satisfy the Pauli exclusion principle, the many-electron wave function is expressed as a Slater determinant \cite{Yu2024}.

The HF process begins with an initial guess for the molecular orbitals. The Fock matrix is then constructed and diagonalized to obtain updated orbitals and their corresponding energies (eigenvalues) \cite{Huang2020}. This iterative cycle continues, with the Fock matrix being reconstructed using the updated orbitals, until the change in total energy between successive iterations falls below a predefined threshold, indicating convergence \cite{Matveeva2023}. Once convergence is achieved, the total electronic energy, representing the GSE, is computed from the final set of orbitals \cite{Crisostomo2022}.

\subsection{Machine Learning}

Machine Learning (ML) is highly effective at learning from data, making it an ideal tool for modeling and understanding small quantum systems. By identifying patterns and relationships within these systems, ML can extrapolate findings to larger, more complex, and computationally challenging systems. In this approach, ML is trained on datasets composed of small molecules, feature descriptors, and quantum properties, enabling it to predict these properties for larger molecules. We propose this as a complementary method to directly solving complex equations.

\section{IMPLEMENTATION}
\subsection{Workflow}

The methodology for this project followed a structured approach. First, an optimal pipeline for computing GSEs of computationally feasible molecules was developed. Strategies tested included the HF approximation, Hamiltonian encoding for molecular fragments, and the parallel application of VQE and QPE to identify the most effective method. Results were then compared against existing literature to evaluate the approach's feasibility.

\begin{figure}[H]
    \centering
    \includegraphics[width=.95\textwidth]{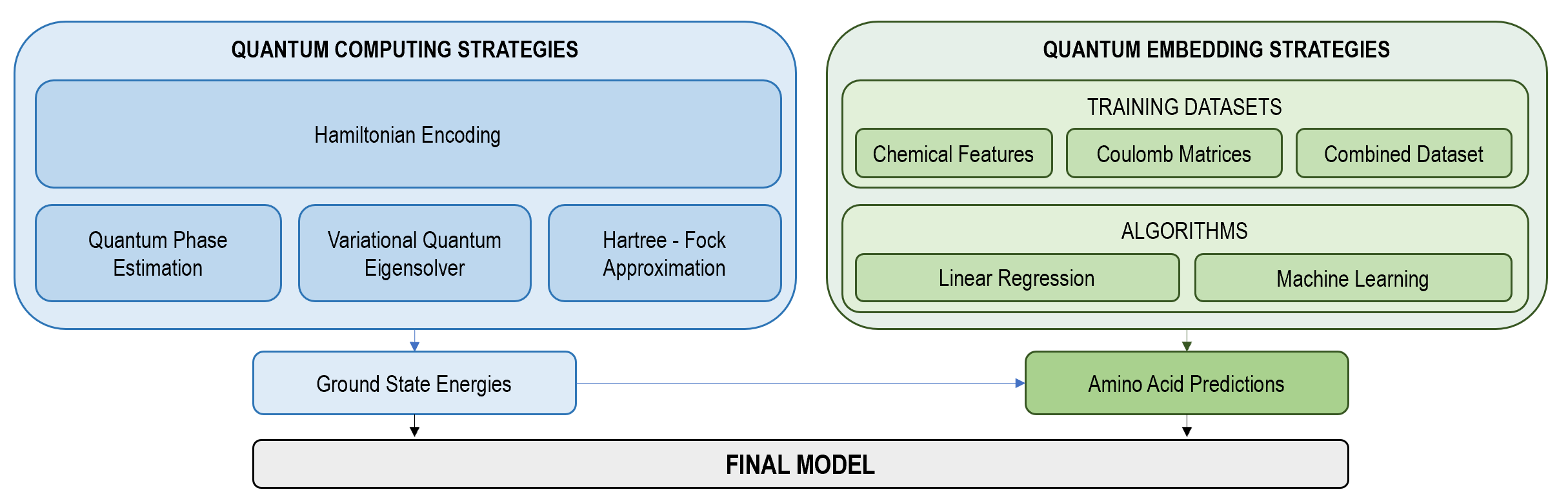}  
    \caption{Pipeline implemented. The left section explores QC strategies for GSE computation, while the right section outlines the development and testing of ML models using training datasets, culminating in the evaluation of the final model.}
    \label{fig:1}
\end{figure}

Subsequently, multiple datasets were created to train ML algorithms to predict the GSEs of larger molecules. To identify the most predictive features, three distinct training datasets were designed: one containing chemical features from the PubChem database \cite{pubchem}, another including Coulomb matrices derived from the $3$D atomic coordinates reported in PubChem, and a third "combined" dataset comprising both sets of features. These datasets were used to train three different ML models to determine the best-performing combination. Additional datasets containing the same molecular descriptors were created for the 20 essential amino acids to test generalization, enabling the evaluation of the trained ML models. SHAP values for the top five predictive features were calculated during training to understand their impact on energy predictions. Finally, the performance of each model was assessed by calculating the Relative Error (RE) between predictions and reference energies from the literature. For a visual overview of the pipeline, see Figure \ref{fig:1}.

\subsection{Computation of Ground State Energies}

First, the Hamiltonian of each generated fragment was encoded to compute the GSE using the Jordan-Wigner transformation. For a two-electron molecule, the parameterized function minimized by the VQE algorithm is represented as Equation \ref{eq:vqe} \cite{kirby2021}.

\begin{equation}
\ket{\Psi(\theta)} = \cos\left(\frac{\theta}{2}\right) \ket{1100} - \sin\left(\frac{\theta}{2}\right) \ket{0011},
\label{eq:vqe}
\end{equation}

Where \(|\Psi(\theta)\rangle\) is the quantum state parameterized by \(\theta\), and \(|1100\rangle\) and \(|0011\rangle\) are basis states of a multi-qubit system.

The first strategy tested was VQE. The electronic Hamiltonian of the molecule was constructed, requiring details such as the number of electrons, orbitals, multiplicity, and molecular charge for accuracy. The simulation device, a double-excitation circuit, was then defined. A gradient-descent optimizer was set to minimize the cost function, initializing \(\theta\) at $0$ (Hartree-Fock state). The algorithm iteratively adjusted \(\theta\) and calculated the corresponding energy until a minimum was reached. This pipeline was applied to compute the GSE of Glycine (\ce{C2H5NO2}) and its fragments (\ce{H2, O2, NH3+, COOH-}).

\end{multicols}

\begin{multicols}{2}

The second approach implemented was QPE, following the detailed steps outlined in \textit{Section $3.1.1$}. The GSE results were compared with those from the VQE and reported literature.

Lastly, the HF approximation was used as an additional pipeline, employing the Pyscf package \cite{pyscf}. Coordinates and atomic details, along with total spin, charge, and basis set, were input to the \textit{gto()} and \textit{scf()} functions. The total system energy was computed iteratively until convergence was achieved. This method was also used to compute the GSE for the same groups tested in VQE and QPE, enabling a comparative analysis to determine the optimal pipeline in our case. 
The development environment used was Visual Studio Code \cite{vscode} with Python as the programming language. Essential libraries included Pennylane and Pyscf for running quantum simulations \cite{pennylane, pyscf}.

\subsection{Datasets}

Due to the complexity of simulating larger molecules, datasets were developed to train ML algorithms for predicting the GSE of molecules such as amino acids. While previous studies have shown promising results using Coulomb matrices as input features \cite{himmetoglu2016}, datasets incorporating common chemical descriptors needed to be more present. This project aimed to determine if combining chemical descriptors with Coulomb matrices could improve prediction accuracy.

With this objective, three training datasets were designed: one containing only chemical features, another with Coulomb matrices derived from atomic positions, and a third combining both types. Each dataset was used independently to train ML models and assess which combination of descriptors yielded the best GSE predictions.

The chemical features dataset included:

\begin{itemize}
    \item Required qubits for computation.
    \item One-hot encoding for atoms (H, C, N, S, O) present in essential amino acids.
    \item Polar surface area.
    \item Molecular complexity.
    \item Logarithm of the octanol-water partition coefficient (logP).
    \item Number of non-hydrogen atoms.
    \item Number of hydrogen bond donors.
    \item Number of hydrogen bond acceptors.
    \item Number of rotatable bonds.
    \item Molecular weight.
\end{itemize}

The Coulomb Matrices dataset included the upper triangular part of the symmetric Coulomb matrices, vectorized and zero-padded to uniform dimensions. These matrices were calculated using equation \eqref{eq:coulomb} and the 3D atomic coordinates from PubChem.

\begin{equation}
C_{IJ} = 
\begin{cases} 
\frac{1}{2} Z_I^{2.4} & \text{if } I = J \\
\frac{Z_I Z_J}{|R_I - R_J|} & \text{if } I \neq J,
\end{cases}
\label{eq:coulomb}
\end{equation}

Where \(C_{IJ}\) represents the matrix element between atoms \(I\) and \(J\), \(Z_I\) and \(Z_J\) are their atomic numbers, and \(|R_I - R_J|\) is the distance between them.

Labels for prediction were the GSEs from the QM7 database, validated with existing literature \cite{hoja2021, cccbdb}. Inclusion criteria for molecules in the datasets were:

\begin{itemize}
    \item Composed of $H$, $C$, $N$, $S$, or $O$ to reflect the functional groups in the 20 essential amino acids.
    \item Energies documented in the QM7 database \cite{hoja2021}.
\end{itemize}

Three test datasets were created using the same criteria but focused on the $20$ essential amino acids. These datasets evaluated the GSE prediction performance of trained ML models. The objective was to identify the optimal combination of data and algorithms for amino acids with potential applications to protein-level predictions.

\subsection{Machine Learning}

The selected algorithms and their tunable hyperparameters were the following:

\begin{itemize}
    \item \textbf{Extreme Gradient Boosting (XGB)} \cite{himmetoglu2016, scikit-learn-GBR}: learning rate, number of estimators, maximum tree depth, subsample, and minimum child weight.
    \item \textbf{Light Gradient Boosting Machine (LGBM)} \cite{Alghushairy2023, Ke2017}: maximum leaves per tree, number of estimators, maximum tree depth, subsample, learning rate, and minimum child weight.
    \item \textbf{Random Forest (RF)} \cite{scikit-learn}: maximum tree depth, number of estimators, minimum samples for node splits, maximum number of features.
\end{itemize}

\section{RESULTS}
\subsection{Ground State Energies}

\begin{figure}[H]
  \centering
  \includegraphics[width=0.3\textwidth]{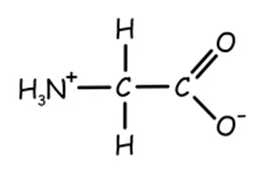}
  \caption{Molecular structure of Glycine, the smallest amino acid.}
  \label{fig:gli}
\end{figure}

The developed pipelines were first tested for computing computationally feasible molecules. Glycine, the smallest of the 20 essential amino acids (shown in Figure \ref{fig:gli}), was chosen for these simulations. The simulations for Glycine and its component groups required significant computation time, with the largest taking over $15$ hours on a system with a 13\textsuperscript{th} Gen Intel(R) Core(TM) i7-13700H processor at 2.40 GHz, 32 GB of RAM, and a 64-bit Windows 11 Pro operating system.

Table \ref{tab:table_comparision_design_index} presents the optimized GSEs obtained using VQE, QPE, and HF, compared to the true GSEs reported in the literature \cite{cccbdb}. The QPE method outperformed both HF and VQE, achieving a mean relative error (\% RE) of $0.18 \pm 0.14\%$, compared to $0.44 \pm 0.66\%$ for HF and $4.19 \pm 3.56\%$ for VQE.

\end{multicols}
\begin{table}[H]
    \centering
    \renewcommand{\arraystretch}{1.5}
    \caption{Optimized GSEs obtained with VQE, QPE, and HF compared to the true GSEs from the literature \cite{cccbdb}.}
    \label{tab:table_comparision_design_index}
    \begin{tabular}{c|c|c|c|c}
        \hline \rowcolor{white} \textbf{Molecule} & \textbf{True GSE [Ha]} & \textbf{VQE [Ha]} & \textbf{QPE [Ha]} & \textbf{HF [Ha]} \\
        \hline \rowcolor{white} \hline
        \textbf{H\textsubscript{2}} & $-1.14$ & $-1.14$ & $-1.14$ & $-1.12$ \\ \hline \rowcolor{white}
        \textbf{O\textsubscript{2}} & $-147.63$ & $-143.98$ & $-147.92$ & $-147.63$ \\ \hline \rowcolor{white}
        \textbf{NH\textsubscript{3}\textsuperscript{+}} & $-55.46$ & $-53.11$ & $-55.62$ & $-55.45$ \\ \hline \rowcolor{white}
        \textbf{COOH\textsuperscript{-}} & $-185.62$ & $-179.03$ & $-184.95$ & $-185.33$ \\ \hline \rowcolor{white}
        \textbf{C\textsubscript{2}H\textsubscript{5}NO\textsubscript{2}} & $-279.12$ & $-249.24$ & $-279.03$ & $-278.35$ \\ \hline \rowcolor{white}
    \end{tabular}
\end{table}
\begin{multicols}{2}

\end{multicols}
\begin{table}[H]
    \centering
    \renewcommand{\arraystretch}{1.5}
    \caption{Description of chemical features in the training and test datasets (amino acids).}
    \label{tab:4}
    \resizebox{0.95\columnwidth}{!}{
    \begin{tabular}{m{5cm}|c|m{5cm}|c}
    \rowcolor{white}
    \multicolumn{2}{c|}{\textbf{TRAINING DATABASE}} & \multicolumn{2}{c}{\textbf{AMINO ACID DATABASE}} \\
    \hline
    \rowcolor{white}
    \textbf{Feature} & \textbf{Mean ± Std [Min, Max]} & \textbf{Feature} & \textbf{Mean ± Std [Min, Max]} \\
    \hline
    \rowcolor{white}
    Molecular weight & $87.12 \pm 21.17 \, [26.04, 124.16]$ & Molecular weight & $136.90 \pm 30.86 \, [75.07, 204.22]$ \\
    \hline
    \rowcolor{white}
    Polar area & $27.52 \pm 19.42 \, [0.00, 71.80]$ & Polar area & $81.76 \pm 20.11 \, [49.30, 128.00]$ \\
    \hline
    \rowcolor{white}
    Complexity & $67.81 \pm 46.26 \, [0.00, 213.00]$ & Complexity & $120.27 \pm 47.26 \, [42.90, 245.00]$ \\
    \hline
    \rowcolor{white}
    Octanol-water partition coefficient & $0.61 \pm 1.17 \, [-3.20, 4.40]$ & Octanol-water partition coefficient & $-2.65 \pm 0.80 \, [-4.20, -1.10]$ \\
    \hline
    \rowcolor{white}
    Number of non-hydrogen atoms & $6.00 \pm 1.25 \, [2.00, 7.00]$ & Number of non-hydrogen atoms & $9.35 \pm 2.41 \, [5.00, 15.00]$ \\
    \hline
    \rowcolor{white}
    Number of hydrogen bond donors & $0.50 \pm 0.71 \, [0.00, 3.00]$ & Number of hydrogen bond donors & $2.65 \pm 0.59 \, [2.00, 4.00]$ \\
    \hline
    \rowcolor{white}
    Number of hydrogen bond acceptors & $1.38 \pm 0.97 \, [0.00, 4.00]$ & Number of hydrogen bond acceptors & $3.70 \pm 0.66 \, [3.00, 5.00]$ \\
    \hline
    \rowcolor{white}
    Number of rotatable bonds & $0.80 \pm 1.13 \, [0.00, 4.00]$ & Number of rotatable bonds & $2.85 \pm 1.18 \, [1.00, 5.00]$ \\
    \hline
    \rowcolor{white}
    Number of qubits required & $46.86 \pm 11.00 \, [14.00, 66.00]$ & Number of qubits required & $76.30 \pm 15.18 \, [40.00, 108.00]$ \\
    \hline
    \rowcolor{white}
    \end{tabular}}
\end{table}
\begin{multicols}{2}

\subsection{Datasets}

Six datasets were developed: three for training and three for testing. The training datasets included $168$ molecules composed of H, C, N, S, and O, encompassing the functional groups found in amino acids. The test datasets included $20$ molecules corresponding to the essential amino acids. Table \ref{tab:4} details the chemical features used in each dataset.

\subsection{Machine Learning}
\subsubsection{Training and Model Selection}

\begin{figure}[H]
    \centering
    \includegraphics[width=.9\textwidth]{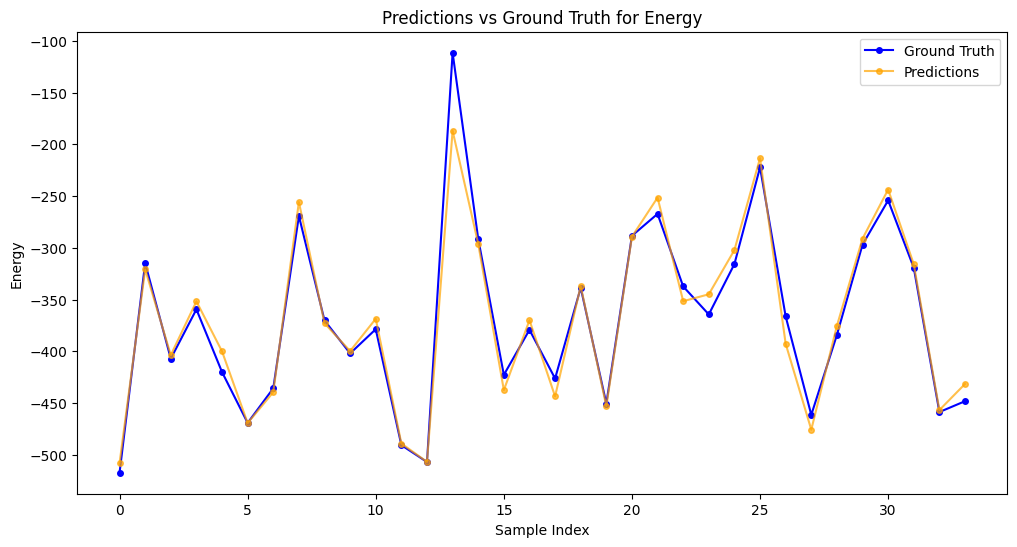}  
    \caption{Predictions during training of the XGB model on the chemical dataset, with energy values on the y-axis and sample indices on the x-axis. The blue line represents the ground truth, and the yellow shows model predictions.}
    \label{fig:7}
\end{figure}

Three ML models — XGB, LGBM, and RF — were trained on the chemical features, Coulomb matrices, and combined datasets. The \% Relative Error (RE) was used to evaluate performance, and the best hyperparameters were stored. Figure \ref{fig:7} illustrates an example from the training phase, showing that the models effectively learned from the data.

\begin{figure}[H]
    \centering
    \includegraphics[width=.9\textwidth]{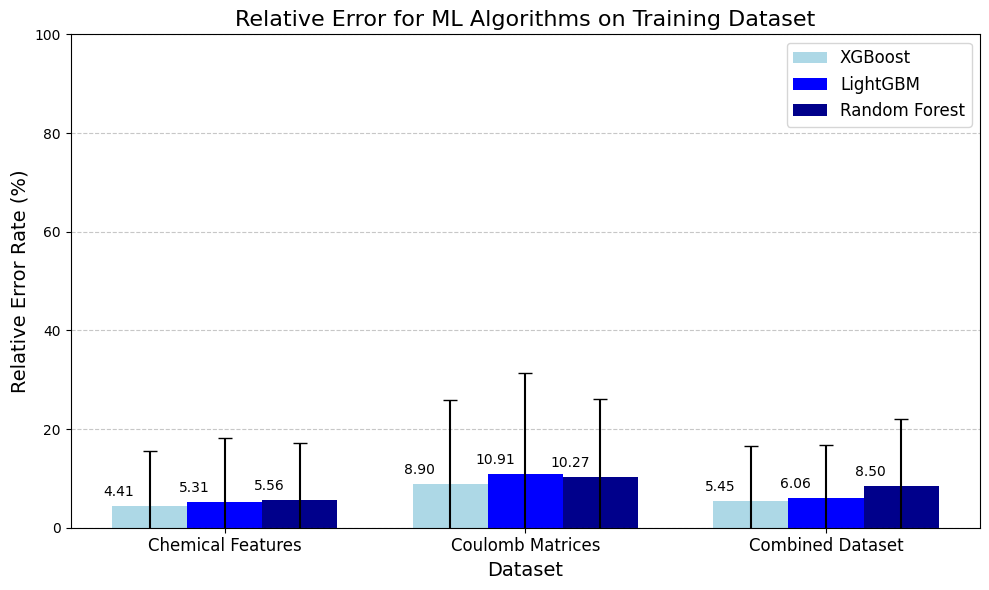}  
    \caption{\% RE for each ML algorithm (different shades of blue) using different feature sets to predict GSE (x-axis). XGB, trained on the chemical features dataset, achieved the best performance.}
    \label{fig:8}
\end{figure}

Figure \ref{fig:8} shows the training results. XGB outperformed other models on the chemical features dataset, achieving a mean RE of $4.41 \pm 11.18\%$. However, performance was lower for the Coulomb matrices dataset, with XGB obtaining an RE of $8.9 \pm 17.01\%$, and all models exhibiting higher errors. LGBM showed the best performance for the combined dataset with an RE of $6.06 \pm 10.79\%$. Thus, the final model expected to perform best during testing was XGB applied to the chemical features dataset with its optimized hyperparameters (n\_estimators: $500$, learning\_rate: $0.05$, max\_depth: $5$, subsample: $0.6$, min\_child\_weight: $5$).

\begin{figure}[H]
    \centering
    \includegraphics[width=0.9\textwidth]{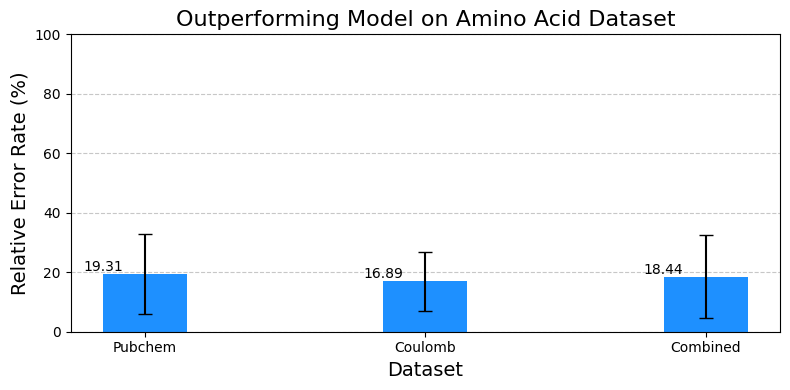}  
    \caption{\% RE for the predictions made by the top-performing algorithms when applied to amino acid features.}
    \label{fig:10}
\end{figure}

\subsubsection{Prediction of Amino Acid Energies}

The best-performing algorithms for each feature set were used to predict the GSEs of the $20$ essential amino acids. For example, if XGB performed best on the combined dataset during training, XBG and its hyperparameters were used to predict amino acid energies with in the combined dataset (see Figure \ref{fig:7} for details on the top model for each case). Figure \ref{fig:10} shows the \% RE values for GSE predictions for the $20$ essential amino acids using chemical, Coulomb, and combined features.

\subsubsection{Explainability}

\begin{figure}[H]
    \centering
    \includegraphics[width=.9\textwidth]{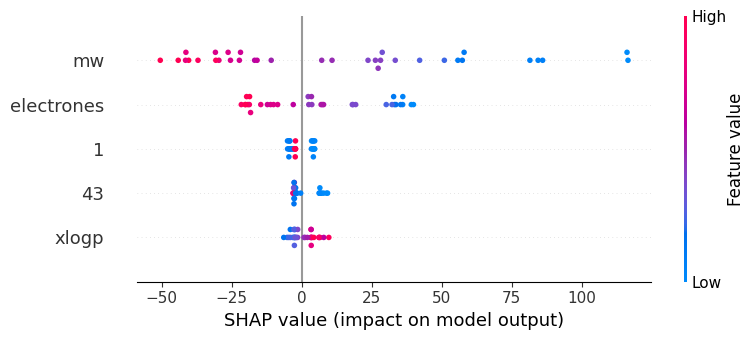}  
    \caption{SHAP values indicating the most predictive chemical features.}
    \label{fig:13}
\end{figure}

Incorporating explainability into high-performing ML algorithms provides valuable insights into their decision-making process. The Shapley Additive exPlanations (SHAP) method was employed to rank features by importance, highlighting their contributions to the model's predictions. Figure \ref{fig:13} shows the SHAP plot for the XGB model trained on the combined dataset. The most significant chemical descriptors identified were molecular weight, number of electrons, partition coefficient, and two specific positions from the Coulomb matrices, 1 and 43.

\begin{figure}[H]
    \centering
    \includegraphics[width=0.9\linewidth]{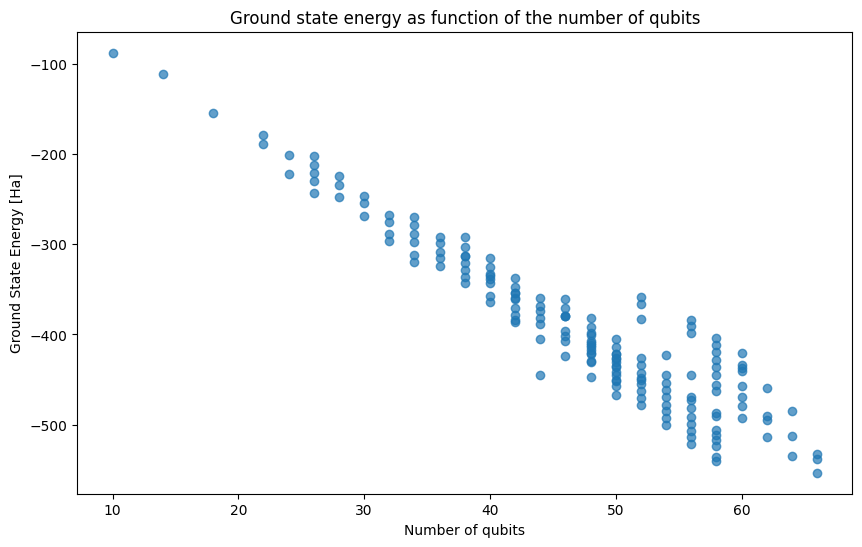}
    \caption{GSEs of $168$ molecules as a function of the number of qubits required for their quantum simulations.}
    \label{fig:14}
\end{figure}

Further analysis revealed that some features showed strong relationships with GSE. The most notable example, shown in Figure \ref{fig:14}, illustrates the apptoximate linear relationship between the GSE of the $168$ molecules and the number of electrons, suggesting that a simple linear model could predict this energy.

\subsubsection{Linear Regression}

\begin{figure}[H]
    \centering
    \includegraphics[width=.8\textwidth]{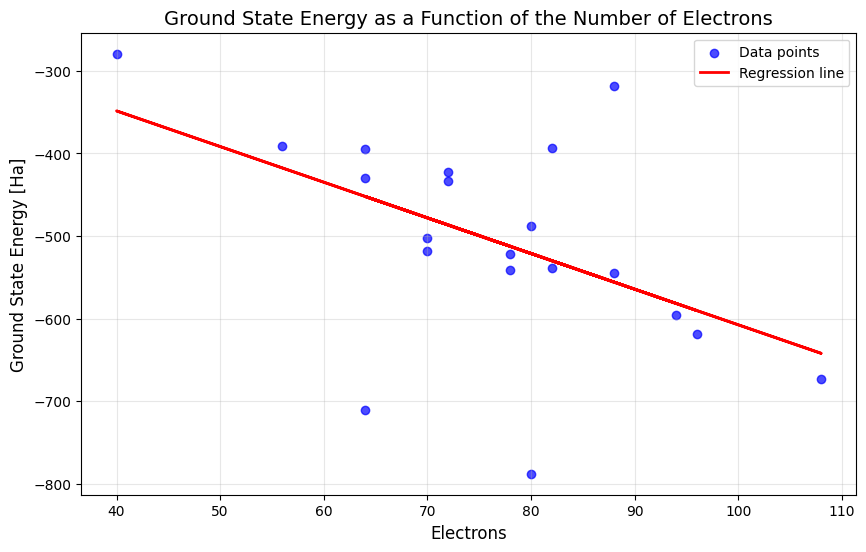}  
    \caption{Linear regression model for predicting the GSE of the 20 essential amino acids. The plot shows the actual GSE as a function of the number of electrons (blue) and the applied linear model (red).}
    \label{fig:16}
\end{figure}

While ML algorithms were designed to predict GSEs using various feature sets, a strong correlation between GSE and the number of electrons was observed. This prompted using a linear regression model for prediction, as shown in Figure \ref{fig:16} and described by equation \eqref{eq:linear}.

\begin{equation}
\text{GSE} = -4.32 \times \text{Qubits} - 163.51
\label{eq:linear}
\end{equation}

Applying this linear model resulted in a \% RE of $15.01 \pm 7.17\%$.

\section{DISCUSSION}

This study was motivated by the potential of QC to improve molecular characterization for drug discovery. Three main objectives were defined: (1) to identify the optimal pipeline for estimating GSE of small, computationally feasible molecules, (2) to develop robust training datasets for training ML models, and (3) to prove if this ML were able to make accurate predictions on larger molecules.

First, implementing QC techniques such as QPE, VQE, and the HF approximation yielded promising results for small molecular fragments, aligning well with findings in the literature \cite{hartree1928wave, kirby2021, peruzzo2014variational}. However, for larger molecules, these methods exhibited limitations in terms of both accuracy and efficiency. VQE achieved a mean relative error (RE\%) of \(4.19 \pm 3.56\%\), while QPE and HF demonstrated significantly lower errors. These differences are expected, as VQE relies on classical optimization algorithms that can be computationally expensive and prone to converging to local minima, especially in complex systems.

In contrast, QPE and HF outperformed VQE, achieving mean RE\% values of and \(0.18 \pm 0.14\%\) and \(0.44 \pm 0.66\%\), respectively, even for larger molecules such as glycine. However, HF's accuracy may decrease for larger systems, suggesting that post-HF methods, such as MP2 or Density Functional Theory, could serve as potential alternatives \cite{gupta2024toward}. Furthermore, QPE is susceptible to circuit noise, so future research could explore combining QPE with techniques like qubitization \cite{low2019}.

Regarding the second and third objectives, integrating ML offered a complementary solution to QC limitations. Three training datasets—chemical features, Coulomb matrices, and a combined set—were developed to determine the optimal molecular descriptors. XGB showed the best performance on chemical features (\% RE of $4.41 \pm 11.18\%$), followed by RF and LGBM (\% RE of $5.56 \pm 11.66\%$ and $5.32 \pm 12.87\%$, respectively). These findings underscore the importance of chemical descriptors in enhancing model accuracy, which is a novel addition as no comparable datasets with these features were found. Chemical descriptors are also more accessible to compute and widely available via databases like PubChem \cite{pubchem}. XGB trained on chemical features, with optimized hyperparameters, emerged as the model expected to perform best when applied to amino acids (n\_estimators: $500$, learning\_rate: $0.05$, max\_depth: $5$, subsample: $0.6$, min\_child\_weight: $5$).

While ML models showed promise during training, their application to the $20$ essential amino acids revealed higher RE values, highlighting challenges in generalization for larger, complex molecules. The combined dataset improved prediction accuracy in the training process, but had slightly higher RE\% in the test set. However, the limited training sample size ($168$ molecules) likely constrained model performance. With this, it seems like probably PubChem features increase accuracy when the molecules are similar but may decrease a bit the generalization capability of the model.

Despite these challenges, the models showed potential for extrapolation to larger molecules. Differences among amino acids in terms of properties and energy levels could explain the varying prediction accuracies, consistent with previous studies that reported acceptable results for more minor, non-polar molecules using fragmentation-based GSE approaches \cite{li2005}.

Additionally, SHAP analysis provided additional insights into the importance of features. For XGB trained on combined features, molecular weight and the number of electrons required were the most influential, suggesting a relationship between molecule size and energy. The partition coefficient, was also significant, reflecting the importance on the nature of the molecule. The remaining features were two elements of the Coulomb matrix, interestingly, the coefficients in positions 1 and 43 which were not null in almost all molecules.

A notable finding was the nearly linear relationship between GSE and the number of electrones of the molecule. Applying a simple linear regression model for GSE prediction resulted in a mean \% RE of $-4.32 \pm -163.51\%$, outperforming some ML models. This result should be interpreted cautiously, as the linear model was based on only $20$ data points, raising potential overfitting concerns. Future research should evaluate this model's applicability to larger systems, such as small proteins.

\section{CONCLUSION}

This study highlights the potential of integrating quantum computing (QC) and machine learning (ML) to enhance molecular characterization and drug discovery. Motivated by the question, \textit{Can deeper insights into subatomic behavior enhance molecular characterization?}, we explored strategies to overcome computational limitations. Our findings show that QC techniques like VQE, QPE, and the Hartree-Fock (HF) approximation are effective for small molecules, with QPE achieving the besr mean relative error (RE) of $0.18 \pm 0.14\%$. However, scaling these methods to larger biomolecules is still challenged by computational noise and limitations, suggesting the future use of post-HF methods for better accuracy.

ML proved to be a valuable complement, extending QC capabilities to larger systems. Developing diverse datasets with chemical descriptors and Coulomb matrices underscored the importance of feature selection, with XGB models achieving the lowest RE of $4.41 \pm 11.18\%$ on chemical descriptors. However, applying these models to the $20$ essential amino acids revealed limitations, highlighting the need for larger datasets to improve accuracy.

A notable finding was the nearly linear correlation between GSE and the number of electrons of the molecule, pointing to potential predictive shortcuts. While a linear regression model showed promise, further evaluation is needed for larger molecules.

In conclusion, combining QC and ML offers a path forward for more precise molecular characterization. Continued development of these methods to handle complex systems could significantly impact computational chemistry and biomedicine, where accurate $3$D molecular modeling is vital. This work lays the groundwork for future research to fully leverage subatomic insights.

\section*{Code}
The code to reproduce the figures and explore additional settings is available in the following GitHub repository: https://github.com/laiacoronas/ML-vs-quantum-algorithms.

\bibliographystyle{unsrt}
\bibliography{references}
\end{multicols}
\end{document}